%% file: gsp_paper.tex
\documentclass[11pt,journal,onecolumn,draftclsnof
oot]{IEEEtran}
\usepackage{amssymb,amsmath,amsthm}
\usepackage{paralist}

\input{packages}
\input{mysymbols.sty}

\begin{document}

 \title{Graph Signal Processing: History, Development, Impact, and Outlook}

\author{Geert Leus, Antonio G. Marques,
Jos\'{e} M. F. Moura, Antonio Ortega,
David I Shuman
\thanks{Geert Leus is with the Fac.~of Electrical Engineering, Mathematics and Computer Science, Delft University of Technology, Delft, The Netherlands, g.j.t.leus@tudelft.nl}
\thanks{Antonio G. Marques is with the Dept.~of Signal Theory and Communications, King Juan Carlos University, Madrid, Spain, antonio.garcia.marques@urjc.es}
\thanks{Jos\'{e} M.~F.~Moura is with the Dept.~of Electrical and Computer Engineering, Carnegie Mellon University, Pittsburgh, Pennsylvania, USA, moura@ece.cmu.edu}
\thanks{Antonio Ortega is with the Dept.~of Electrical and Computer Engineering, University of Southern California, Los Angeles, California, USA, aortega@usc.edu}
\thanks{David I Shuman is with Franklin W. Olin College of Engineering, Needham, Massachusetts, USA, dshuman@olin.edu}
}
\date{}

\maketitle




Signal processing (SP) excels at analyzing, processing, and inferring information defined over \emph{regular} (first continuous, later discrete) domains such as time or space. 
Indeed, the last 75 years have shown how SP has 
made an impact in 
areas such as communications, acoustics, sensing, image processing, and control, to name a few. With the digitalization of the modern world and the increasing pervasiveness of data-collection mechanisms, information of interest in current applications oftentimes arises in non-Euclidean, irregular domains. Graph signal processing (GSP) generalizes SP tasks to signals living on non-Euclidean domains whose structure can be captured by a weighted graph. Graphs are versatile, able to capture irregular interactions, easy to interpret, and endowed with a corpus of mathematical results, rendering them natural candidates to serve as the basis for a theory of processing signals in more irregular domains. 

The term ``graph signal processing'' was coined a decade ago in the seminal works \cite{sgwt,EmergingFieldGSP,SandryMouraSPG_TSP13,SandryMouraSPG_TSP14Freq}.
Since these papers were published, GSP-related problems have drawn significant attention, not only within the SP community \cite{Stankovic2020book,ortega2022introduction} but also in machine learning venues, where research in graph-based learning has increased significantly \cite{bronstein2017geometric}. Graph signals are well-suited to model measurements/information/data associated with (indexed by) a set where: (i) the elements of the set belong to the same class (regions of the cerebral cortex, members of a social network, weather stations across a continent), (ii) there exists a relation (physical or functional) of proximity, influence or association among the different elements of that set, and (iii) the strength of such a relation among the pairs of elements is not homogeneous.
In some scenarios, the supporting graph is a physical, technological, social, information or biological network where the links can be explicitly observed. In many other cases, the graph is implicit, capturing some notion of dependence or similarity across nodes, and the links must be inferred from the data itself. 
As a result, GSP is a broad framework that  encompasses and extends classical SP methods, tools, and algorithms to application domains of the modern technological world, 
including 
social, transportation, communication and brain networks; recommender systems; financial engineering;  distributed control; and learning, just to name a few. While the theory and application domains of GSP continue to expand, GSP has become a technology with wide use. It is a research domain pursued by a broad community, the subject of not only many journal and conference articles, but also of textbooks \cite{Stankovic2020book,ortega2022introduction}, special issues of different journals, symposia, workshops, and special sessions at ICASSP and other SP conferences.

In this article, we provide an overview of the evolution of GSP, from its origins to the challenges ahead. The first half is devoted to reviewing the history of GSP and explaining how it gave rise to an encompassing framework that shares multiple similarities with SP, and especially digital signal processing (DSP). A key message 
is that GSP has been critical 
to develop novel and technically sound tools, theory, and algorithms that, by leveraging analogies with and the insights of DSP, provide new ways to analyze, process, and learn from graph signals. 
In the second half, we shift focus to review the impact of GSP on other disciplines. First, we look at the use of GSP in data science problems, including graph learning and graph-based deep learning. Second, we discuss the impact of GSP on applications, including neuroscience and image and video processing. We conclude with a  brief discussion of the emerging and future directions of GSP. 

\section{The Early Roots}


The roots of GSP can be traced to  algebraic and spectral graph theory, harmonic analysis, numerical linear algebra, and specific  applications of these ideas to areas such as 
data representations for high-dimensional data, pattern recognition, (fast) transforms, image processing, computer graphics, statistical physics, partial differential equations, semi-supervised learning, and neuroscience. Algebraic graph theory \cite{algebraicgraphtheory2001} dates back to the 1700s, and spectral graph theory \cite{spectralgraphtheory1980,Chung:96} dates back to the mid-1900s. They study mathematical properties of graphs and link the graph structure to the spectrum (eigenvalues and eigenvectors) of matrices related to the graph. However, they generally did not consider potential signals that could be living on the graph. 

In the late 1990s and early 2000s, graph-based methods for analyzing and processing data became more popular -- independently -- in a number of disciplines, including computer graphics (e.g., \cite{taubin1995signal}), image processing (e.g, \cite{WuLeahy93,elmoataz}), graphical models in Bayesian statistics (e.g.,  \cite{kolaczyk2009book,wainwright2008graphical}), dimensionality reduction (e.g., \cite{Roweis:00,Belkin:03}) and semi-supervised learning (e.g., \cite{SmolaKondor2003}) within machine learning, and neuroscience (e.g., \cite{brain2006}, detailed history included in \cite{booksporns2010networks}). 
For example, in computer graphics, Taubin utilized graph Laplacian eigenvectors to perform surface smoothing by applying a low pass graph filter to functions defined on polyhedral surfaces \cite{taubin1995signal}, and later used similar ideas to compress polygonal meshes. 
In image processing, weighted graphs can be defined with edges being a function of pixel distance and intensity differences. Such semi-local and non-local graphs were exploited for denoising (bilateral filtering), image smoothing, and image segmentation (e.g., \cite{bilateral1998,elmoataz,WuLeahy93}). 
Graphical models \cite{wainwright2008graphical} -- and in particular undirected graphical models, also referred to as Markov random fields -- model data as a family of random variables (the vertices), with the graph edges capturing their probabilistic dependencies. Through the graph, these models sparsely encode complex probability distributions in high-dimensional spaces. Graphical models have been widely used in Bayesian statistics and Bayesian probabilistic approaches, kernel regression methods, statistical learning, and statistical mechanics \cite{koller2009probabilistic}. We return to semi-supervised learning and neuroscience and their connections with GSP in Sec. \ref{Se:ssl} and Sec. \ref{Se:neuro}, respectively.

Also in the late 1990s, \cite{watts1998strogatz,barabasialbert-science1999}
introduced two new models for random networks (graphs) to model the  structure of complex engineered systems, going well beyond the classical Erd\"os-R\'enyi random graphs; real world large networked systems exhibit small world characteristics \cite{watts1998strogatz} and scale free degree distributions  \cite{barabasialbert-science1999}. This led to a flurry of activity, usually referred to as \textit{Network Science} \cite{national2006network}, concerned with analyzing and designing complex systems like telecommunication, power grid, and large scale infrastructure networks \cite{newman2018networks}. While the central focus
of network science was on properties of the network and its nodes (e.g.,  centralities, shortest paths, and clustering coefficients),
network science researchers also leveraged graphs to explore the dynamics of processes such as percolation, traffic flows, synchronization, and epidemic spread \cite[Part V]{newman2018networks}, often adopting mean field approximations. For example, in the investigation of the susceptible-infected-susceptible (SIS) epidemiological model in scale-free graphs in \cite{pastor2001epidemic}, each vertex can be seen as having a 0/1 (susceptible/infected) signal residing on it. Advancements in network science have certainly informed the subsequent development of GSP. 


In parallel, a stream of new methods for analyzing data on graphs were investigated. These methods tried specifically to combine (i) intuition and dictionary constructions for performing computational harmonic analysis on data on Euclidean domains with (ii) generalizable ways to incorporate the structure of the underlying graph into the data transforms. For example, one of the first general wavelet constructions for signals on graphs was the spatial wavelet transform of \cite{graphwavelets2003}, which was defined directly in the vertex domain. In the seminal work \cite{COIFMAN2006}, diffusion wavelets were constructed by (i) creating a multiresolution of approximation spaces, each spanned by graph signals generated by diffusing a unit of energy outwards from each vertex for a fixed amount amount of time; and (ii) computing orthogonal diffusion wavelets to serve as basis functions for the detail spaces that are the sequential orthogonal complements of the approximation spaces. Spectral graph wavelets \cite{sgwt} traded off the orthogonality of diffusion wavelets for a simpler generative method for each wavelet atom: define a pattern in the graph spectral domain and localize that pattern to be centered at each vertex of the graph. Meanwhile, the algebraic signal processing approach \cite{Pueschel:08a} showed that classical SP can be captured by a triplet defined by a shift operator. Different shifts lead to different SP models and different Fourier transforms. In particular, \cite{Pueschel:08b} showed that a shift based on Chebyshev polynomials, appropriate for lattice models like in images, leads to standard block transforms such as the discrete cosine and  Karhunen–Lo\`{e}ve transforms (DCT and KLT) that can be understood as Fourier transforms on certain graphs. Numerous other types of multiresolution transforms and dictionaries for data residing on graphs, trees, and compact manifolds were investigated in the subsequent few years. These included lifting and pyramid transforms, graph filter banks, tight spectral frames, vertex-frequency transforms that generalized the classical short-time Fourier transform, and learned dictionaries (see \cite{ortega2018graph,9244195} for a more complete literature review and list of references). GSP arose from these different fields, coalescing
multiple perspectives into a common framework and set of ideas. In the last decade, this unifying framework has evolved into a full-fledged theory and technology.

\section{The Theoretical Underpinnings}

Ten years ago, \cite{sgwt,EmergingFieldGSP,SandryMouraSPG_TSP13,SandryMouraSPG_TSP14Freq} introduced the field of GSP and established many of its foundations. Remarkably, these works approached the problem from two different perspectives. Inspired by graph theory and harmonic analysis, the authors of \cite{sgwt,EmergingFieldGSP} use the graph Laplacian as the core of their theory, naturally generalizing concepts such as frequencies and filter banks to the graph domain. Differently, the authors of \cite{SandryMouraSPG_TSP13,SandryMouraSPG_TSP14Freq} follow an algebraic approach, under which the multiplication of a graph signal by the adjacency matrix of the supporting graph yields the most basic operation of shift for a graph signal. Based on this simple operation, more advanced tools such as filtering, graph Fourier transform, graph frequency, or total variation can be generalized to the vertex and spectral graph domains. Rather than being considered as competing approaches, these works brought complementary views and tools and, jointly, contributed to increase the attention to the field. After introducing some common notation, this section reviews these two approaches, and then explains how they were merged into an integrated framework that facilitated drawing links with classical SP and propelled the growth of GSP.

\subsection{Basic definitions and notational conventions}

The goal in GSP is to leverage SP and graph theory tools to 
analyze and process signals defined over a network domain, with notable examples including technological, social, gene, brain, knowledge, financial, marketing, and blog networks. In these setups, graphs are used to both index the data and to represent relations/similarities/dependencies among the locations of the data. We denote the underlying weighted graph by $\mathcal{G}=(\mathcal{V},\mathcal{E},\omega)$, where $\mathcal{V}:=\{1,...,N\}$ denotes the set of $N$ graph vertices, $\mathcal{E}\subset \mathcal{V}\times \mathcal{V}$ denotes the set of graph edges, and $\omega: \mathcal{E} \rightarrow \mathbb{R}$ is a weight function that assigns a real-valued weight to each edge, with a higher edge weight representing a stronger similarity or dependency between the two vertices connected by that edge. A graph with edge weights all equal to 1 is called \emph{unweighted}.
A \emph{graph signal} contains information associated with each vertex of the graph. For simplicity, we focus our discussion on scalar, real-valued graph signals (each signal is a mapping from $\mathcal{V}$ to $\mathbb{R}$), but the values associated with each node could be discrete, complex, or even vectors (e.g., when multiple features per node are observed). Each scalar, real-valued graph signal can equivalently be represented as an $N$-dimensional vector $\mathbf{x}:=[x_1,...,x_N]^\top$, with $x_i$ (also written sometimes as $[\mathbf{x}]_i$) representing the value of the signal at vertex $i$. An example of a graph signal is shown in Fig. \ref{fig:frequency_domain}.


To gain some intuition, consider the problem of studying Twitter patterns. Assume that we have $N$ Twitter users, each vertex $i\in \mathcal{V}$ represents a user~$i$, and each edge $e=(i,j)\in \mathcal{E}$ captures that two users~$i$ and~$j$ follow each other. The data, $x_i$, indexed by node~$i$ could, e.g., be the number of tweets that user~$i$ tweeted in a given time interval. In a second application, to understand traffic flow in cities, we can examine the number of pick-ups of for-hire-vehicles (FHV) (e.g., taxis, Uber or Lyft cars, etc) over a given time period. The graph $\mathcal{G}$ can be the city road map, with the vertices $i\in \mathcal{V}$ representing intersections and the edges $e\in \mathcal{E}$ representing road segments between intersections. The data $x_i$ at each vertex~$i$ might, e.g., be the number of pick-ups close to that intersection over the time period of interest. The graphs~$\mathcal{G}$ in such real-world applications can be modeled as undirected (if $(i,j) \in \mathcal{E}$, then $(j,i) \in \mathcal{E}$), or directed (e.g., to capture one-way streets).

Classical SP signals such as audio and image signals that reside on Euclidean domains can also be viewed as graph signals. Consider for instance finite-length discrete-time one-dimensional signals, e.g., the~$N$ vertices of the graph are the time instances~$i=0,...,N-1$, with $N$ being the window length. Since the signal value $x_{i+1}$ at time $i+1$ is usually closely related to the signal value $x_i$ at the preceding time, there is a directed edge from vertex~$i$ to vertex $i+1$. At $i=N-1$, there are different options for the boundary conditions; here, we consider the periodic boundary condition, which means that the time instant ``next'' to the terminal instant $N-1$ is $i=0$. The resulting ``time graph'' is then a directed cycle $\mathcal{G}_{dc}$ (see Fig. 2). By similar reasoning, vertices in the image graph represent the pixels, and because the image brightness or color $x_{ij}$ at pixel $(i,j)$ is usually highly related to the brightness or colors of its four neighboring pixels, there are undirected edges from $(i,j)$ to its neighboring pixels. The corresponding graph
is then an undirected $2D$-lattice.


At the core of GSP are $N \times N$ matrices that encode the graph's topology. The most prominent are (i) the adjacency matrix $\mathbf{A}$, whose $(i,j)$-entry is the edge weight $\omega((i,j))$ if $(i,j) \in {\mathcal{E}}$ and 0 otherwise; and (ii) the combinatorial (or non-normalized) graph Laplacian $\mathbf{L}:=\mathbf{D}-\mathbf{A}$, where $\mathbf{D}=\mathrm{diag}(\mathbf{A}\mathbf{1})$ is the diagonal matrix of vertex degrees (sums of the weights of the edges adjacent to each vertex) and $\mathbf{1}$ is an $N \times 1$ vector of all ones; and (iii) the normalized graph Laplacian $\mathbf{L}_{\mbox{norm}}:=\mathbf{D}^{-\frac{1}{2}}\mathbf{L}\mathbf{D}^{-\frac{1}{2}}$. We elaborate on the role of these matrices in the next section.


\subsection{The spectral approach for GSP} 

Classical Fourier analysis of a one-dimensional signal decomposes the signal into a linear combination of complex exponential functions (continuous or discrete) at different frequencies, with increasing frequencies corresponding to higher rates of oscillation and basis functions that are less smooth. The spectral approach to GSP \cite{sgwt,EmergingFieldGSP} generalizes this classical Fourier analysis by writing graph signals as linear combinations of a basis of graph signals with the property that the basis vectors can be (roughly) ordered according to how fast they oscillate across the graph, or, related, how smooth they are with respect to the underlying graph structure. By ``smooth'' in this context, we mean that the values of the graph signal at each pair of neighboring vertices are similar. 

The operator that captures this notion of smoothness with respect to the underlying (undirected) graph is the graph Laplacian $\mathbf{L}$. It is a discrete difference operator, since we have 
\[ [ {\bf L}{\bf x} ]_i = \sum_{j=1}^N A_{i,j} (x_i -x_j) = \sum_{j \in \mathcal{N}_i} A_{i,j} (x_i -x_j),\]
where $\mathcal{N}_i$ is the neighborhood of node $i$. Since ${\bf L}$ is a real symmetric matrix, it has a set of orthonormal eigenvectors $\{ {\bf v}_\ell \}_{\ell=0}^{N-1}$ and a set of real non-negative eigenvalues $\{ \lambda_\ell \}_{\ell=0}^{N-1}$. Assuming a connected graph, it can further be shown that there is only one eigenvalue zero, e.g., $\lambda_0=0$, with corresponding eigenvector ${\bf v}_0={\bf 1}/\sqrt{N}$.
In matrix form we obtain
${\bf L} = {\bf V} \text{diag}({\boldsymbol \lambda}) {\bf V}^\top$, 
with ${\bf V} = [ {\bf v}_0, \dots, {\bf v}_{N-1} ]$ and ${\boldsymbol \lambda} = [\lambda_0, \dots, \lambda_{N-1}]^T$.  

Importantly, the graph Laplacian can also be viewed as a graph extension of the time-domain Laplacian operator $\frac{\partial^2}{\partial t^2}$. Just as the  one-dimensional complex exponentials -- the eigenfunctions of the time-domain Laplacian operator -- capture a notion of frequency, we can interpret the graph Laplacian eigenvectors as graph frequency vectors, with the associated graph Laplacian eigenvalues capturing a notion of the rate of oscillation \cite{EmergingFieldGSP}. 

\begin{figure}
    \centering
    \includegraphics[width=0.8\textwidth]{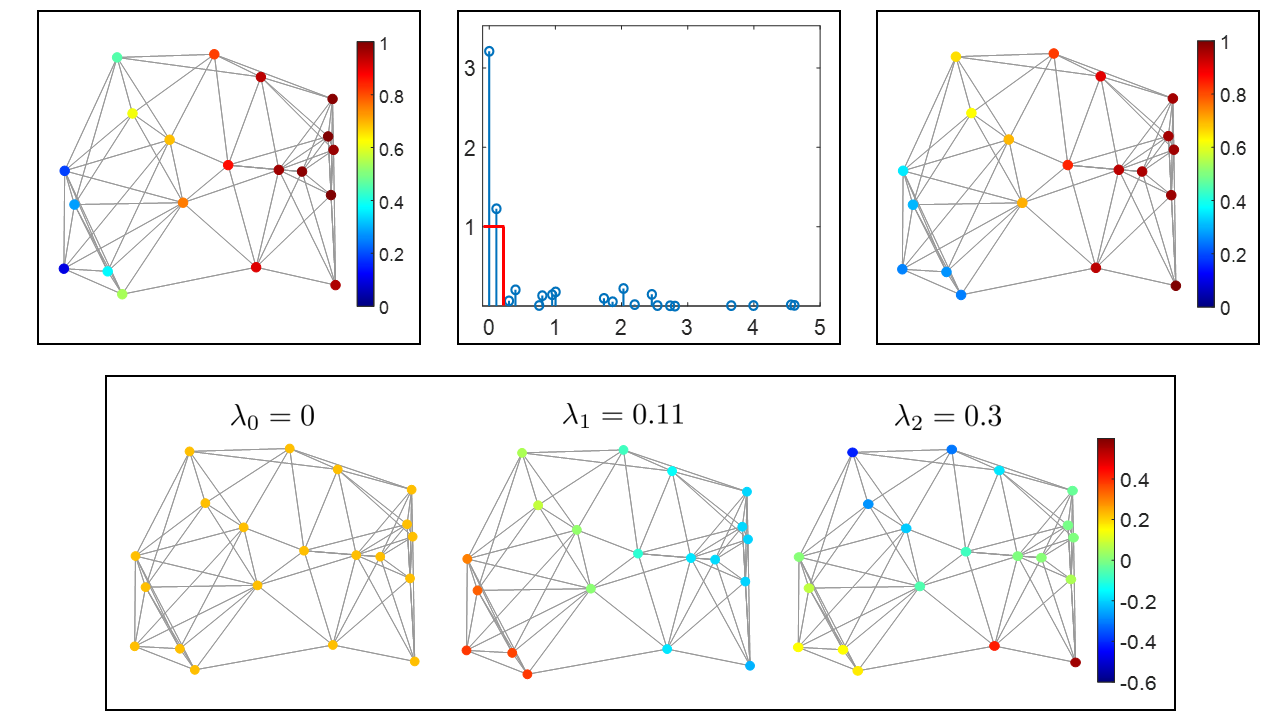}
    \caption{Top left: An example of a graph with a color-coded graph signal on top; Top middle: The signal in the graph frequency domain and in red the frequency response of a potential low-pass graph filter; Top right: The filtered graph signal; Bottom: The first three eigenvectors of the graph Laplacian ordered with decreasing smoothness (increasing eigenvalue).}
    \label{fig:frequency_domain}
\end{figure}

The Laplacian operator can also be used to introduce a measure of smoothness for a graph signal ${\bf x}$; namely, the graph Laplacian quadratic form
\begin{equation}\label{Eq:quadform}
{\bf x}^\top {\bf L} {\bf x} = \sum_{(i,j) \in \mathcal{E}} A_{i,j} (x_i - x_j)^2, 
\end{equation}
which penalizes large differences between signal values at strongly connected vertices. Because ${\bf v}_\ell^\top {\bf L} {\bf v}_\ell = \lambda_\ell$, it is then clear from \eqref{Eq:quadform} that the larger the graph frequency $\lambda_\ell$, the less smooth (or more variable) the graph Laplacian eigenvector ${\bf v}_\ell$. So 
with the indexing convention $0= \lambda_0 < \lambda_1 \leq \dots \leq \lambda_{N-1}$, the graph frequency vectors $\{ {\bf v}_\ell \}_{\ell=0}^{N-1}$ are ordered according to increasing variability (see Fig.~\ref{fig:frequency_domain}).
Using the Laplacian eigenvectors as the basis,
we can now define a graph Fourier transform (GFT) as ${\bf V}^\top$. It transforms a graph signal ${\bf x}$ into its frequency components as $\hat{\bf x} = {\bf V}^\top {\bf x}$.




Graph filters can then be interpreted as operators that modify the different frequency components of a signal ${\bf x}$ individually. That is, the graph filter operation can be represented in the graph Fourier domain  by $\ccalH: \reals \rightarrow \reals$, so that $[\hat{\bby}]_\ell=\ccalH(\lambda_\ell)[\hat{\bbx}]_\ell$. In most cases, the spectral function $\ccalH$ (oftentimes referred to as a kernel) is set to a pre-specified analytical form (typically parametric) that promotes certain properties in the output signals (e.g., rectangular kernels promote smoothness and remove noise; see Fig.~\ref{fig:frequency_domain}). However, non-parametric approaches can also be used. Equally important, \cite{EmergingFieldGSP} also illustrates how graph filters can be used 
to interpolate missing values and to design signal dictionaries whose atoms concentrate their energy around a few frequencies or vertices, highlighting their relevance in a number of applications.  

\subsection{The algebraic approach for GSP} 

In classical SP, convolution is a key building block present in most algorithms, including filtering and sampling, and interpolation. 
In defining convolution and filtering, the time shift -- the unit delay that transforms a signal into a delayed version of itself -- plays a critical role. The outputs of linear time invariant filters are weighted linear combinations of delayed versions of the input. Similarly, the DFT can be understood as the transformation that diagonalizes every linear time invariant filter and provides an alternative description for signals and filters. 

In extending these ideas to GSP,
two key contributions of \cite{SandryMouraSPG_TSP13,SandryMouraSPG_TSP14Freq}  are: (i) highlighting the relevance of defining a ``graph-aware'' operator that plays the role of the ``most basic operation'' to be performed on a signal $\mathbf{x}$ defined over a graph $\mathcal{G}$; and (ii) setting this operation as $\mathbf{A}\mathbf{x}$, i.e., the multiplication of the graph signal $\mathbf{x}$ by the adjacency matrix $\mathbf{A}$ of $\mathcal{G}$. The motivation for the latter choice is twofold. First, $\mathbf{A}$ is a simple (parsimonious and linear) operator that combines the values of $\mathbf{x}$ in a manner that accounts for the local connectivity of $\mathcal{G}$. Second, when particularized to time-varying signals defined over the directed cycle $\mathcal{G}_{dc}$, using $\mathbf{A}_{dc}\mathbf{x}$ is equivalent to the classical unit delay; i.e., $[\mathbf{A}_{dc}\mathbf{x}]_{i+1}=[\mathbf{x}]_{i}$. 

How can this basic graph-aware operator be leveraged to design (i) linear graph filters that are applied to a graph signal to generate another graph signal, and (ii) linear transforms that provide an alternative representation for a graph signal? In classical SP,\footnote{\label{ftn:DSP-finite}Because graphs are finite, we consider DSP with finite signals, and, for simplicity, with periodic signal extensions.} the basic non-trivial operation applied to a signal is the unit delay (time shift); in other words, the simplest filter is the time shift filter $z^{-1}$. Generic linear filters are then polynomials of this basic operator of the form $p(z)=p_0+p_1 z^{-1}+\cdots + p_1 z^{-(N-1)}$, with $z^{-l}$ being the consecutive application of the operator $z^{-1}$ to a time signal $l$ times. DSP polynomial filters are shift invariant in the sense that $z^{-1}\cdot p(z)=p(z)\cdot z^{-1}$.


Hence, to address the first question, \cite{SandryMouraSPG_TSP13} sets the simplest signal operation in GSP as the multiplication by the adjacency matrix~${\bf A}$ and, subsequently, defines graph filters as (matrix) polynomials of the form $p({\bf A})=p_0 {\bf I}_N+p_1 {\bf A}+\cdots + p_1 {\bf A}^{(N-1)}$. It is easy to see that polynomial filters are $\bbA$ invariant, in the sense that $ {\bf A} \cdot p({\bf A})=p({\bf A})\cdot {\bf A}$. Apart from the theoretical motivation, the polynomial definition exhibits a number of advantages. When applied to a graph signal $\bbx$, the operation $\bbA\bbx$ can be understood as a local linear combination of the signal values at one-hop neighbors. Similarly, $\bbA^2\bbx$ is a local linear combination of $\bbA\bbx$, reaching values that are in the 2-hop neighborhood. From this point of view, a graph filter $p(\bbA)$ represented by a polynomial of order $L$ is mixing values that are at most $L$ hops away, with the polynomial coefficients $\{p_l\}_{l=0}^L$ representing the strength given to each of the neighborhoods. Another advantage is that, if $\bbA$ is set to the $\bbA_{dc}$ (the graph representing the support of classical time signals) the graph polynomial definition $p(\bbA_{dc})$ reduces to the classical time shift definition $p(z^{-1})$, so that graph filters become linear time invariant filters. 

\begin{figure}
    \centering
    \includegraphics[width=0.6\textwidth]{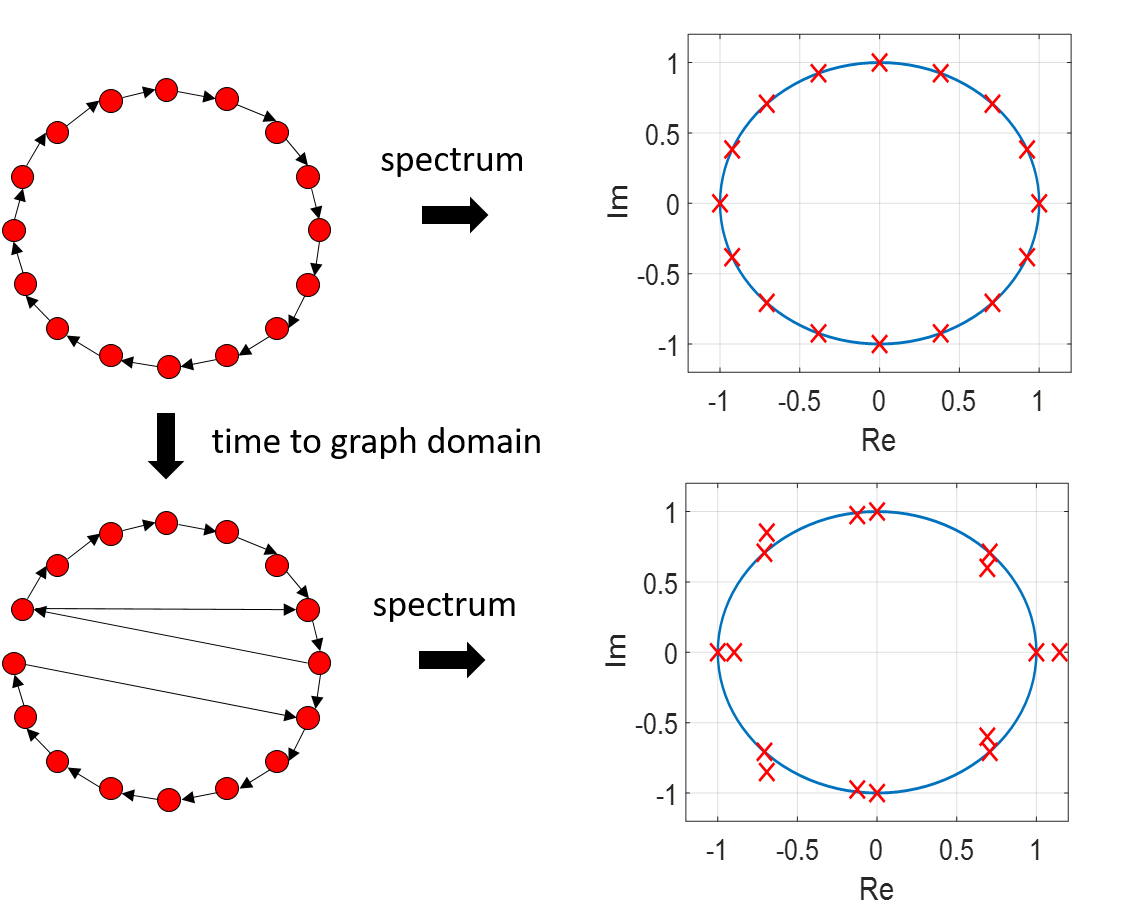}
    \caption{Left: From the directed cycle representing the time domain to a general graph; Right: Eigenvalues (spectrum) of the related adjacency matrices.}
    \label{fig:time_to_graph}
\end{figure}

To address the second question,  \cite{SandryMouraSPG_TSP13} defines the graph Fourier transform (GFT) as the linear transform that diagonalizes these graph filters of the form $p(\bbA)$.  
Letting $\bbA=\bbV\diag(\bblambda)\bbV^{-1}$ be the eigendecomposition of the (possibly directed) adjacency matrix $\bbA$, 
then $p(\bbA)=p(\bbV\diag(\bblambda)\bbV^{-1})=\bbV(p(\diag(\bblambda))\bbV^{-1}$ (note that we use $\bbV^{-1}$ now instead of $\bbV^\top$ because the eigenvectors are not necessarily orthonormal as for the Laplacian). In other words, matrix polynomials can be understood as operators that transform the input by (i) multiplying it by the matrix $\bbV^{-1}$, (ii) applying an orthogonal operator $p(\diag(\bblambda))$, and (iii) transforming the result back to the vertex domain with a multiplication by $\bbV$. The GFT of a graph signal and the signal spectral representation is then set as the multiplication by $\bbV^{-1}$ and the frequency response of a filter is found by calculating $p(\diag(\bblambda))$, similarly as described in the previous section on the spectral approach. From the GFT of the signal, common SP concepts can now be defined in GSP \cite{SandryMouraSPG_TSP14Freq} including ordering of graph frequencies, low and high graph frequencies, or design of low- and high-pass graph filters.
Fig.~\ref{fig:time_to_graph} shows the generalization of the time domain to a more general graph domain. Applications in \cite{SandryMouraSPG_TSP13} to data prediction, graph signal compression, data classification, and customer behavior prediction for service providers and in \cite{SandryMouraSPG_TSP14Freq} to filter design and malfunction detection in sensor networks show the breadth of application domains.

\subsection{The benefits of a joint framework}
While having different origins, the approaches in \cite{sgwt,EmergingFieldGSP} and \cite{SandryMouraSPG_TSP13,SandryMouraSPG_TSP14Freq} bring complementary perspectives. The work in \cite{sgwt,EmergingFieldGSP} relies on the graph Laplacian to capture the structure of $\mathcal{G}$, uses its eigendecomposition to characterize graph signals and define filtering operations, and draws clear links with existing graph-based techniques in a number of applications. In \cite{SandryMouraSPG_TSP13,SandryMouraSPG_TSP14Freq}, the focus is on the shift operation in the vertex domain, postulating the use of the adjacency matrix as the building block to design GSP algorithms, and unveiling a number of similarities with classical SP. Although some early works mixed features of \cite{sgwt,EmergingFieldGSP,SandryMouraSPG_TSP13,SandryMouraSPG_TSP14Freq} (e.g., the  use of polynomials based on the Laplacian matrix), the publication of these four papers and related works led to the emergence of works that combine both approaches under a common framework. One way to do so is to define a generic ``graph-shift operator'' (GSO) that plays a dual role: (i) it can be viewed as the most basic operation applied to a graph signal, and (ii) it codifies the structure of the graph in a more generic way than $\bbL$ or $\bbA$, so that it can be used to tackle a broader range of setups. Under this framework, the linear GSO $\mathbf{S}\in\reals^{N\times N}$ has been set to different adjacency matrices (e.g., one-hop, two-hops), different graph Laplacians (e.g., combinatorial, normalized, random walk), the precision matrix of a Gaussian Markov random field, or even combinations of those. Based on the eigendecomposition of this operator, given by $\mathbf{S}=\mathbf{V}\mathrm{diag}(\boldsymbol{\lambda}) \mathbf{V}^{-1}$, linear graph filtering can be equivalently understood as an operator that is linear and orthogonal (diagonal) in the frequency domain defined by $\bbV^{-1}$, or as the multiplication by a matrix that is a linear combination of successive applications (powers) of the GSO $\bbS$: 
\begin{align}\label{eq:graph_filters_vertex_freq}
\mathbf{H}(\bbS)=\mathbf{V} \mathrm{diag}(\widehat{\mathbf{h}}) \mathbf{V}^{-1}~~\text{or}~~\mathbf{H}(\bbS)=\sum_{l=0}^{N-1}h_l\mathbf{S}^l,
\end{align}
where the notation $\mathbf{H}(\mathbf{S})$ is used to emphasize the dependence on the GSO $\bbS$. The first definition in \eqref{eq:graph_filters_vertex_freq} focuses on the frequency domain, with the filter parameters being the $N$-dimensional frequency response $\widehat{\mathbf{h}}=[\widehat{h}_0,...,\widehat{h}_{N-1}]^\top$. The second definition in \eqref{eq:graph_filters_vertex_freq} focuses on the vertex domain, with the parameters of the filter being the $N$ filter taps $\mathbf{h}=[h_0,...,h_{N-1}]^\top$. Although we focus on degree $N-1$ polynomials, thanks to the Cayley–Hamilton Theorem, the definition in \eqref{eq:graph_filters_vertex_freq} can represent a matrix polynomial of any degree \cite{SandryMouraSPG_TSP13}. With these models at hand, the literature promptly addressed tasks such as prediction, classification, compression, filter identification, and filter design in graph/network contexts \cite{SandryMouraSPG_TSP13,segarra2017optimal,isufi2016autoregressive}.
The particular solution obtained for any of these tasks depends on the GSO at hand, as well as the assumptions on the graph filter. For example, if the goal is to estimate the graph-based linear mapping from a set of input-output pairs collected in matrices $\bbX=[\bbx_1,...,\bbx_M]$ and $\bbY=[\bby_1,...,\bby_M]$, one requires $M=N$ input-output pairs if no structure is assumed for $\bbH$ and a single $M=1$ pair if one assumes that $\bbH$ is a graph filter. Equally important, defining the counterparts of classical finite impulse response (FIR) and infinite impulse response (IIR) filters as $\mathbf{H}_{FIR}(\bbS)=\sum_{l=0}^{L-1}b_l\mathbf{S}^l$ and $\mathbf{H}_{IIR}(\bbS)=\big(\sum_{l=0}^{L-1}a_l\mathbf{S}^l)^{-1}$, identification from input-output observations is feasible even if only a subset (with cardinality larger than $2L$) of the signal values is observed \cite{isufi2016autoregressive,djuric2018cooperative}. Additionally, using the definitions in \eqref{eq:graph_filters_vertex_freq}, it is not difficult to show that any cascade/parallel/feedback connection of graph filters can also be written as a graph filter, opening the door to make and exploit connections between graph-network processes and classical tools in control \cite{djuric2018cooperative}.

A natural next step is to use \eqref{eq:graph_filters_vertex_freq} to model certain properties of classes of graph signals of interest. To be more specific, consider that we model a graph signal $\bbx\in\reals^N$ from a class of interest as  $\mathbf{x}=\mathbf{H}(\mathbf{S})\mathbf{z}$ with $\bbz$ being a hidden seed signal and $\mathbf{H}(\mathbf{S})$ a generative graph filter that ``transfers'' some of the properties of $\bbS$ to $\bbx$. 
While mathematically simple, modelling graph signals as $\mathbf{x}=\mathbf{H}(\mathbf{S})\mathbf{z}$ has proven to be fruitful. A typical approach is to assume some parsimonious structure on either $\mathbf{z}$, the filter $\mathbf{H}(\mathbf{S})$, or both, and then analyze the impact of those assumptions on the properties of $\mathbf{x}$. Standard assumptions have included $\mathbf{H}(\mathbf{S})$ being a bandlimited graph filter so that $\mathbf{x}$ is graph bandlimited \cite{chen2015discrete}, $\mathbf{H}(\mathbf{S})$ being low pass so that $\mathbf{x}$ is smooth \cite{EmergingFieldGSP,dong2016learning,mateos2019connecting}, $\mathbf{z}$ being a white signal so that $\mathbf{x}$ is graph stationary  \cite{marques2017stationary,perraudin2017stationary,djuric2018cooperative}, or $\mathbf{z}$ being sparse so that $\mathbf{x}$ is a diffused graph signal~\cite{segarra2017blind}, as well as combinations of those. More importantly, the combination of the generative model $\mathbf{x}=\mathbf{H}(\mathbf{S})\mathbf{z}$  and one or more of the previous structural assumptions have been leveraged to successfully generalize a number of estimation and learning tasks to the graph domain. Early examples investigated in the literature included signal denoising, sampling and interpolation, input identification, blind deconvolution, dictionary design, semi-supervised learning, classification, and the generalization of stationarity to graph domains (see, e.g, \cite{djuric2018cooperative,ortega2018graph} for detailed reviews). While covering all of these tasks goes beyond the scope of this paper, we discuss next three illustrative milestones: (i) sampling and interpolation, (ii) source identification and blind deconvolution, and (iii) statistical descriptions of random graph signals. 

We start with a simple sampling and interpolation setup that, due to is practical relevance, received early attention from multiple research groups \cite{Tanaka2020SPM}.
Consider the  sampling set $\ccalM \subseteq \ccalV$ with cardinality $M \leq N$, and define the selection matrix $\bbPhi_\ccalM \in \{0,1\}^{M \times N}$ as the $M$ rows of the $N \times N$ identity matrix indexed by the set $\ccalM$. The sampled signal $\bbx_\ccalM:=\bbPhi_\ccalM\bbx$ collects the values of the graph signal $\bbx$ at the vertex set $\ccalM$. The goal is to use $\bbx_\ccalM$ -- along with $\bbS$ -- to recover $\bbx$, leveraging the structure of the graph. Since the problem is ill posed, we need to assume and enforce some structure on $\bbx$. Two widely adopted approaches are to (i) assume that $\bbx$ is $K$-bandlimited; i.e., it is in the span of the first $K$ eigenvectors of $\bbS$, for some $K<N$, or (ii) assume that the signal $\bbx$ is smooth with respect to the underlying graph, which can be generically modelled as the norm of $\|\bbx - \bbH(\bbS)\bbx\|$ being small, where $\bbH(\bbS)$ is a low pass filter that is tuned to promote a particular notion of smoothness. We denote the subspace of $K$-bandlimited signals by $\ccalX(\bbV_K):=\{\bbV_K\boldsymbol{\beta}~\forall~ \boldsymbol{\beta}\in \reals^K\}$.
The statement that $\bbx \in \ccalX(\bbV_K)$ is equivalent to saying that $\bbx$ is generated via a graph filter with $\widehat{\mathbf{h}}=[\mathbf{1}_K^\top,\mathbf{0}_{N-K}^\top]^\top$. 
These two alternative assumptions lead to the following optimization problems for interpolation, respectively:
\begin{equation}\label{Eq:samp_interp}
    \bbx^* = \mathrm{arg min}_{\bbx\in \ccalX(\bbV_K)} \
     \|\bbx_\ccalM-\bbPhi_\ccalM\bbx\|_2^2 \mbox{~~or~~}  \bbx^* = \mathrm{arg min}_{\bbx}\|\bbx_\ccalM-\bbPhi_\ccalM\bbx\|_2^2 + \alpha \|(\bbI-\bbH(\bbS))\bbx\|_2^2,
\end{equation}
with the weight $\alpha$ controlling the trade-off between minimizing the smoothness of $\bbx^*$ and how similar $\bbx^*$ and $\bbx$ are for the nodes in $\ccalM$. For bandlimited signals, if $M\geq K$ and $(\bbPhi_\ccalM\bbV_K)$ is full rank, the signal $\bbx$ can be identified from its samples $\bbx_\ccalM$ via $\bbx=\bbV_K(\bbPhi_\ccalM\bbV_K)^\dagger\bbx_\ccalM$ \cite{chen2015discrete}. While this is also true for time signals, other popular results in classical SP -- such as ideal low-pass filters being the optimal interpolators or regularly spaced sampling being optimal -- do not hold true for the graph domain, due to the lack of regularity in $\ccalG$. Regarding the second optimization problem in \eqref{Eq:samp_interp}, the solution is $\bbx^*=\big(\bbPhi_\ccalM^\top\bbPhi_\ccalM + \alpha(\bbI-\bbH(\bbS)^2\big)^{-1}\bbPhi_\ccalM^\top\bbx_\ccalM$. In this case, we can interpret $\bbPhi_\ccalM^\top\bbx_\ccalM$ as a zero-padded graph signal that is smoothly diffused through the graph by $\big(\bbPhi_\ccalM^\top\bbPhi_\ccalM + \alpha(\bbI-\bbH(\bbS)^2\big)^{-1}$.

Using the model $\mathbf{x}=\mathbf{H}(\mathbf{S})\mathbf{z}$, source identification and blind deconvolution have also been generalized to the graph setting \cite{segarra2017blind}. 
In both, the signal $\bbz$ is assumed to be sparse. For source identification, given a sampled version of $\bbx$, the goal is to identify the locations and non-zero values of $\bbz$, which can be viewed as source nodes whose inputs are diffused throughout the network represented by $\bbS$. For blind deconvolution, the goal is to use $\bbx$ to identify both the sparse input $\bbz$ and the generating filter $\bbH(\bbS)$, with a classical assumption being that the coefficients $\bbh$ are sparse or that the filter has a parsimonious FIR/IIR structure. Inspired by those works, generalizations were also investigated for demixing setups where the aggregation of multiple signals is observed (e.g., the sum of several network processes, each with different sources and dynamics).

Our last example to illustrate the benefits of a common GSP framework is the statistical description of random graph signals. Characterizing random processes is a challenging task even for regular time-varying signals, with stationarity models excelling at finding a sweet spot between practical relevance and analytical tractability. With this is mind, multiple efforts were carried out to generalize the definition of stationarity to graph signals \cite{marques2017stationary,perraudin2017stationary,djuric2018cooperative}. The key step 
was to say that a zero-mean random graph signal $\bbx$ is stationary in $\bbS$ if it can be modeled as $\bbx=\bbH(\bbS)\bbz$, with $\bbz$ being white. This is equivalent to saying that the covariance matrix $\bbC_\bbx=\E{\bbx\bbx^\top}$ can be written as a polynomial of the GSO $\bbS$, illustrating the relation between the underlying graph and the statistical properties of the graph signal, and establishing meaningful links with Gaussian Markov random fields that assume $\bbS=\bbC_\bbx^{-1}$. With this definition, counterparts to concepts and tools such as the power spectral density, the periodogram, the Wiener filter, and autoregressive-moving-average (ARMA) models were developed \cite{marques2017stationary}. These developments provide new ways to design graph-based covariance estimators and denoise graph signals, as well as a rigorous framework to better model, understand, and control random processes residing on a graph.

We close this section by highlighting that, while some instances of the problems discussed had been investigated well before the GSP framework was put forth (e.g., denoising based on smooth priors given by powers of the Laplacian, or source identification based on graph diffusion processes), those early works were mostly disconnected and focused on particular setups. The advent of GSP and the use of a common language and theoretical framework served a number of purposes: (i) facilitating the identification of connections between and differences among existing works; (ii) bringing different research communities together; (iii) enabling the design of more complex processing architectures that use early works as building blocks; (iv) providing a new set of tools for graph signals based on the generalization of classical SP schemes to the graph domain; and (v) aiding the development of novel, theoretically-grounded solutions to graph-based problems that had been solved in a heuristic manner.

\section{The Impact of GSP on Data Science}

GSP has transformed how the SP community deals with irregular geometric data; however, it has also contributed to areas that go beyond SP, having a significant impact on data science related disciplines. To illustrate this, we next review several of the data science problems where GSP-based approaches have made significant contributions. 

 \subsection{Graph learning}

The field of GSP was originally conceived with a given graph ($\ccalG$ or $\bbS$) in mind. Such a graph could originate from a physical network, such as transportation, communication, social or structural brain networks for instance. However, in many applications, the graph is an implicit object that describes relations or levels of association between the variables. In some cases, the links of those graphs can be based on expert domain knowledge (e.g., activation properties in protein-to-protein networks), but in many other cases the graph must be inferred from the data itself.  Examples include
graphs for image processing where the edges are defined based on both pixel distance and intensity differences, a $k$-nearest neighbor ($k$NN) graph for semi-supervised learning where edges connect data points with similar sets of features, or correlation graphs for functional brain networks. In those cases, the problem to solve can be formulated as ``given a collection of $M$ graph signals $\bbX=[\bbx_1,...,\bbx_M]$ 
$\in\reals^{N\times M}$, find an $N\times N$ sparse graph matrix $\bbS$ describing the relations among the nodes of the graph''. Clearly, such a problem is severely ill posed, and models to relate the properties of the graph and the signals are key to address it in a meaningful way. 

Learning a graph from data is a topic on its own, with roots in statistics, network science and machine learning (see~\cite{kolaczyk2009book} and references therein). Initial approaches focused on the information associated with each node separately, so that the existence of the link $(i,j)$ in the graph was decided based only on the $i$th and $j$th row of $\bbX$. Contemporary (more advanced) approaches look at the problem as finding a mapping from $\bbX$ to $\bbS$, with graphical lasso (GL) being the most prominent example. GL is tailored for Gaussian Markov random fields and sets the graph to a sparsified version of the precision matrix, so that $\bbS\approx (\frac{1}{M}\bbX\bbX^\top)^{-1}$~\cite{kolaczyk2009book}. The contribution of GSP to the problem of graph learning \cite{mei_tsp2017,mateos2019connecting,dong2019learning} 
falls into this second class of approaches,
where the more sophisticated (spectral and/or polynomial) relations between the signals and the graph can be fully leveraged. One cluster of early GSP works focused on learning a graph $\bbS$ that made the signals in $\bbX$ smooth with respect to the learnt graph \cite{dong2016learning}. If smoothness is promoted using a Laplacian-based total variation regularizer $\sum_{m=1}^M \bbx_m^\top\bbL \bbx_m$, the formulation leads to a kernel-ridge regression problem with the pseudoinverse of $\bbL$ as the kernel and, meaningful links with GL can be established \cite{dong2019learning}. 
A second set of GSP-based topology inference methods model the data $\bbX$ as resulting from a diffusion process over the sought graph $\bbS$ through a graph filter. The key questions when modeling the observations as $\bbx_m=\bbH(\bbS)\bbz_m$ are then the assumptions (if any) about the diffusing filter $\bbH(\bbS)$ and the input signals $\bbz_m$. Assuming the inputs $\bbz_m$ to be white, which is tantamount to assuming that the signals $\bbx_m$ are stationary in $\bbS$, leads to a model where the covariance (precision) matrix of the observations is a polynomial of the sought GSO $\bbS$, all having the same eigenvectors. This not only provides a common umbrella to several existing graph-learning methods, but also a new (spectral and/or polynomial) way to address graph estimation \cite{segarra2017networktopology,pasdeloup2018}. Indeed, the fact of GSP offering a well understood framework to model graph signals has propelled the investigation of multiple generalizations of the above methods, tackling, e.g., directed graphs, causal structure identification, presence of hidden nodes whose signals are never observed \cite{meimoura-silvar}, dynamic networks, multi-layer graphs, and nonlinear models of interaction. We refer the interested reader to~\cite{mateos2019connecting} and references therein for more details.

\subsection{Network science} 
As discussed above, advancements in network science informed subsequent developments in GSP. It is now also the case that GSP techniques have been used to address network science problems such as clustering and community mining. We mention four examples here. First, in \cite{tremblay2014graph}, spectral graph wavelets are utilized to develop a new fast, multiscale community mining protocol. Second, by graph spectral filtering random graph signals, \cite{tremblay2016compressive} efficiently constructs feature vectors for each vertex in a manner such that the distances between vertices based on these feature vectors resemble the distances based on standard spectral clustering feature vectors \cite{von2007tutorial}. In \cite{tremblay2020approximating}, a detailed account is provided of how that approach and other new sampling and interpolation methods developed for GSP can be used to accelerate spectral clustering by avoiding $k$-means. Third, \cite{donnat2018learning} uses spectral graph wavelets to learn structural embeddings that help identify vertices that have similar structural roles in the network, even though they may be distant in the graph. Finally, \cite{singh2017gft} introduces a new centrality measure based on the GFT.

\subsection{Semi-supervised learning} \label{Se:ssl}

The goal of semi-supervised learning (SSL) is to utilize a combination of labeled and unlabeled data to predict the labels of the unlabeled data points. The labels may be discrete (semi-supervised classification) or continuous (semi-supervised regression). Many of the graph-based SSL methods (e.g., \cite{SmolaKondor2003}) investigated by the machine learning community in the early 2000s constructed an undirected, weighted similarity graph with each vertex representing one data point (either labeled or unlabeled), and then diffused the known labels across the graph to infer the labels at the unlabeled vertices. This approach can also be thought of as compelling the vector of labels to be smooth with respect to the underlying graph. Mathematically, this results in optimization problems with at least two terms: a fitting term that ensures the vector of labels exactly or approximately matches the known labels on the vertices corresponding to the labeled data points, and a regularization term of the form $\mathbf{x}^{\top}\mathbf{H(S)x}$ for some GSO $\mathbf{S}$ and (low pass) graph filter $\mathbf{H(S)}$ that enforces global smoothness of the signal \cite[Sec. III.D]{shuman_distributed_SIPN_2018}.

Rather than enforcing global smoothness of the labels with respect to the underlying graph, another GSP approach to SSL is to encourage the labels to be piecewise-smooth with respect to the graph by modeling them as a sparse linear combination of graph wavelet atoms \cite{gavish2010multiscale,shuman2011semi,ekambaram2013wavelet}. Regularization problems resulting from this approach feature the same fitting term as above, but the additional term in the objective function captures the sparsity prior through the norm (or mixed norm) of the coefficients used to synthesize the labels as a linear combination of the graph wavelets. Finally, in GSP parlance, SSL is intimately related to graph signal interpolation, so that most of the results regarding the sampling and reconstruction of (bandlimited) graph signals, can (and have been) applied to SSL.

\subsection{Graph neural networks}

Neural networks (NN) are non-linear data processing architectures composed of multiple layers, each of which combines (mixes) the inputs linearly via matrix multiplication and then applies a scalar nonlinear function to each of the entries of the output. 
The values of the mixing matrices $\{\bbTheta_{\ell}\}_{\ell=1}^L$ are considered as the parameters of the architecture. To avoid an excess of parameters, a standard approach is to impose some parsimonious structure on the mixing matrices (e.g., Toeplitz, low-rank, sparse), giving rise to different families of NN. Given the success of NN -- and convolutional NN in particular -- in processing regular data such as speech and images, a natural question is how to generalize these architectures to data defined over irregular graph domains. In this context, the ML learning community investigated graph NN (GNN) that incorporate the graph ($\ccalG$ or $\bbS$) into NN architectures in different ways \cite{bronstein2017geometric,Defferrard2016,du2017topology}.
GSP offers a principled way to address this question, postulating that the matrices $\{\bbTheta_{\ell}\}_{\ell=1}^L$ have the form of a graph filter $\{\bbH(\bbS)_{\ell}\}_{\ell=1}^L$. This both offers a 
flexible way to incorporate the graph (with the selection of the GSO $\bbS$  being application dependent) and also provides a range of options for parameterizing the graph filter (e.g., polynomial filters, rational filters, diffusion filters). Similarly, a number of generalizations and novel architectures that leverage GSP have been proposed, including pooling schemes based on sampling over graphs, graph recurrent NN, architectures defined over product graphs, and NN based on graphon filters \cite{marques2020editorial}. GSP has not only provided a common framework to better understand the contributions of and links between many of the existing works, but has also facilitated novel contributions on subjects such as transferability, robustness, or sensitivity with respect to the graph \cite{ruiz2021graph}.  

\subsection{Graph-time processing}

In many applications, a time series -- as opposed to a scalar value -- is observed at each node of the graph $\ccalG$. If the length of each time series is $T$, the data at hand can be arranged in the form of a matrix $\bbX = [ \bbx_1, \dots, \bbx_T ] \in\reals^{N\times T}$, which can be viewed as a collection of $N$ times series (one per node of the graph), as a collection of $T$ graph signals, or as a single signal $\text{vec}(\bbX)\in\reals^{NT}$ that varies across both time and the nodes of the graph $\ccalG$. 
The first approaches to handle time-varying graph signals were based on \emph{product graphs} that combine a graph of the vertices with a graph for the time domain (e.g., a directed cycle graph $\ccalG_{dc}$) to obtain a single larger graph $\ccalG\times\ccalG_{dc}$ with $NT$ nodes~\cite{SandryMouraSPMag,time-vertex-2018}. 
This interpretation allows for the use of standard GSP tools such as the GFT transform and graph filters, with the joint GFT being the Kronecker product of the original GFT $\bbV^{-1}$ and the DFT matrix $\bbF^H$, and the joint GSO some chosen product (e.g., Kronecker, Cartesian, strong) of the respective GSOs. Indeed, the joint spectrum of the time-varying graph signal $\text{vec}(\bbX)$ can be analyzed this way, and joint graph-time filters can be adopted for its denoising or interpolation. In their most general form, those filters need not be separable over the graph and time domains, thereby increasing their modeling and processing potential.

Later, vector autoregressive (VAR) processes were considered for graph-time processing. A VAR models a vector process by expressing the current vector as a matrix-weighted version of past vectors plus some innovation, i.e., $\bbx_t = \sum_{p=1}^P \bbA_p \bbx_{t-p} + \bbe_t$. Considering that the vectors we are handling are graph signals, 
the underlying graph structure can be incorporated in such VAR models in different ways, leading to different GSP extensions.
One direction is to replace the matrix weights by graph filters, i.e., $\bbA_p = \bbH_p (\bbS)$, leading to {\it graph VAR processes}~\cite{graph-VAR-2019}. In such models, the graph filters can be implemented in the graph frequency domain or as a polynomial of the GSO in the vertex domain. Furthermore, causal models have been assumed where the polynomial order of the graph filter cannot be larger than the time delay on which the filter operates~\cite{mei_tsp2017}. 
Another extension of VAR models also considers the interaction between the different nodes of the current vector, i.e., $\bbx_t = \bbA_0 \bbx_t + \sum_{p=1}^P \bbA_p \bbx_{t-p} + \bbe_t$, where $\bbA_0$ has a zero diagonal. It further enforces sparsity on all the matrix weights. In such {\it structural VAR processes}~\cite{ggtopoid2018piee}, the matrix weights can be viewed as graph adjacency matrices that link the current data on a node with past data on the same node as well as with current and past data on neighboring nodes. Extensions to non-linear versions have also been considered.

\section{The Value of GSP in Science and Engineering Applications}

Not surprisingly, GSP methods have been applied to engineering networks where a clear definition of the graph follows from a physical network.  
These include communication networks (e.g., developing distributed schemes to estimate the channels), smart grids, power networks, (e.g., designing distributed resource allocation algorithms for power flow), water networks,  and transportation networks (e.g., developing graph-based architectures to predict traffic delay). Similarly, GSP has also contributed to applications where the network is not explicitly observable but can be inferred from additional information, such as social networks, meteorological prediction, genetics, and financial engineering. While all of the previous examples are meaningful and relevant, we briefly highlight here the two areas with the largest and most consistent GSP activity over the past decade: neuroscience and image and video processing. 

\subsection{Applications to neuroscience} \label{Se:neuro}



Graphs have a long history in neuroscience, since they can be used to represent different relationships and pairwise connections between regions of the brain, taking each region to be a vertex \cite{booksporns2010networks}. An anatomical brain graph captures structural connections between the regions, as measured, e.g., via fiber tracts in white matter captured through diffusion magnetic resonance imaging. A functional brain graph, on the other hand, aims to capture pairwise interdependencies between activity that is measured in the different brain regions. Identifying the functional brain graph has been 
studied extensively for different reasons and with different modalities, the most common of which is functional magnetic resonance imaging (fMRI). 
Often, such studies also involve the estimation of dynamic graphs~\cite{TV_brain_graphs_Montana,dynamic_brain_graph_VDV}. 
During a sequence of task and rest periods, it has for instance been shown that the on-task and off-task 
functional brain graphs differ substantially~\cite{TV_brain_graphs_Montana}. 
Recent work also demonstrates that dynamics in the functional brain graph even exist during resting state fMRI, with meaningful 
correlations with electroencephalograph (EEG), demographic, and behavioral data~\cite{dynamic_brain_graph_VDV}. 

Interestingly, most graph-based approaches in neuroscience consist of first identifying a brain graph and then using graph-theoretical and network-science tools to analyze its properties. From this point of view, GSP tools can be (and have been) leveraged for learning brain graphs \cite{brain_ProcIEEE}. However, GSP really shines when it comes 
to analyzing how the measured activity pattern -- the brain signal -- behaves in relation to a brain graph (either anatomical or functional, related to one or multiple subjects). In other words, 
GSP provides a technology to merge the brain function, contained in the brain signal, with the brain graph (see~\cite{brain_ProcIEEE} and references therein).
Specifically, the GFT has been used 
to analyze cognitive behavior. 
For example, \cite{alignment_signal_brain} shows that 
that there is a relation 
between the energy of the high frequency content of an fMRI signal and the attention switching ability of an individual. There is further 
research that states that, when learning a task, the correlations between 
the learning rate and the energies of the low/high
frequency content of an fMRI signal change with the exposure time; i.e., they depend on how familiar we are with the task~\cite{GFT_brain_AR}. Another study confirms the brain regions that are affected by multiple concussions by comparing the 
differences in energy of the high and low graph frequency components between athletes with a history of multiple concussions and healthy controls~\cite{fMRI_concussions}.
In addition to the GFT, graph wavelets and Slepians have been used to reveal localized frequency content in the brain~\cite{brain_ProcIEEE}, and graph filters have been used as diffusion operators to model disease progression in dementia. 
While these results demonstrate
the potential GSP has for neuroscience, we believe this pairing is still in its infancy and there is plenty of room for exploration. 

\subsection{Applications to image and video processing}

As noted earlier, widely used techniques in image and video processing, including transforms such as the DCT 
and the KLT
, segmentation methods, 
and image filtering 
can be interpreted from a GSP perspective \cite{cheung2018graph}. 
In recent years, the emergence of a broader understanding of GSP has led to a further evolution of how graph-based approaches are used for image processing. As an example, while the DCT 
or the asymmetric discrete sine transform (ADST)  are formed by the eigenvectors of path graphs with equal edge weigths, extensions have been proposed where graph edges with lower weights can be introduced in between pixels corresponding to image contours \cite{hu2014multiresolution}.
In these approaches, as in input-dependent image filtering \cite{Milanfar2013}, the image structure is first analyzed (e.g., contours detected) and then transforms adapted to the image characteristics are selected, with the choice of transform sent as side information. 

A particularly promising application of GSP methods is to point cloud processing and compression. Each point in a point cloud is defined by its coordinates in 3D space and has associated with it an attribute (e.g., color or reflectance). While points are in a Euclidean domain, their positions, on the surfaces of the objects in the scene, are irregular and make it natural to develop a graph-based processing approach. Transforms have been proposed that leverage or are closely related to the GFT of a point cloud graph \cite{zhang2014point,dequeiroz2016compression}. These methods are fundamental algorithms for geometry-based point cloud compression \cite{schwarz2018emerging}. Additionally, point cloud processing has become a major application domain for graph machine learning, with applications in areas such as denoising \cite{valsesia2020deep}.

\section{The Future Ahead}
The focus of this article has been on reviewing the early results and growth of GSP, with an eye not only on the SP community but also on the applications and data science problems that have benefited from GSP. 
We close by discussing some of the emerging directions and open problems that we believe will shape the future of the field. 

One emerging area in the field of GSP is \textbf{dynamic graphs}, more specially, how to estimate them and how to process time-varying graph signals residing on them. Graphs are rarely static; think for instance about social networks with new users or changing connections, or functional brain networks determined by a specific task that is carried out at a particular time instant. As a result, GSP tools, theory, and algorithms need to be extended to such scenarios. There is already quite some work on graph topology identification for dynamic graphs, but most of these methods link consecutive graphs in the cost function, making the problems computationally challenging~\cite{ggtopoid2018piee,mateos2019connecting}. Adaptive methods (of the correction-only or prediction-correction type) try to tackle this issue, but tracking rates are still low. 
Processing signals residing on time-varying graphs has not been studied in depth, and this is clearly an area where many opportunities arise.

Extending GSP to \textbf{higher-order graphs} is another important future direction. Some applications are characterized by a graph domain where more than two nodes can interact; think for instance about a co-authorship network where groups of co-authors that collaborated on a paper are linked together, or about movie graphs in recommender systems, where movies starring the same actor form a group. Such graphs where an edge can join more than two nodes are called higher-order graphs. Popular abstractions of higher-order graphs are  \emph{simplicial complexes} and \emph{cell complexes}. A simplicial/cell complex is a collection of subsets of the set of nodes satisfying certain properties. Whereas in a simplicial complex the subsets satisfy the subset inclusion property (e.g., there need to be links between each pair of the three co-authors of a paper), in a cell complex, they do not. However, both types of complexes share a similar recursive relationship between the higher-order Laplacians, leading to a hierarchical processing architecture that can process node signals over edges, edge signals over triangles/polygons (for a simplicial/cell complex), and so on. 
A less restrictive representation of a higher order graph is a \emph{hypergraph} ${\cal H}=({\cal V},{\cal E},\omega)$, where $\omega$ is a function that assigns a weight to each hyperedge in ${\cal E}$. Hyperedges can connect more than two vertices in ${\cal V}$. 
Some recent overviews on higher-order networks, with focuses on GSP and network science, respectively,
can be found in~\cite{higher-order-Schaub2021,BATTISTON20201}. There are still many open issues in higher order GSP, including the exploration of connections to adjacent fields such as topological data analysis and computational geometry.
Many other open problems -- extending GSP to include uncertainty in the signals and graphs, design of exact and approximate Bayesian (recursive) estimators able to track variations across nodes and time, developing GSP models for categorical data, generalizing GSP results to continuous manifold (geometric) data, incorporating GSP tools into reinforcement learning and spatio-temporal control, etc. -- are also expected to play important roles in the future of the discipline. If the first years of GSP combined theoretical developments with practical applications by placing a stronger focus on the former, we expect that the coming years will see an increased emphasis on applications, along with important efforts on learning and statistical schemes.


{\small
\bibliographystyle{ieeetr}
\bibliography{gsp_paper}
}

\end{document}

%% file: packages.tex
\usepackage[T1]{fontenc}
\usepackage[utf8]{inputenc}
\usepackage{amsfonts,amsmath}
\usepackage{lmodern}
\usepackage[hidelinks]{hyperref}
\usepackage{bookmark}
\usepackage{color}
\usepackage{blindtext}  
\usepackage{fancyhdr}
\usepackage{titlesec}
\usepackage{etoolbox}

\patchcmd{\chapter}{\thispagestyle{plain}}{\thispagestyle{empty}}{}{}   

\usepackage[figuresright]{rotating}
\usepackage{caption}
\usepackage{subcaption}
\graphicspath{{figs/}}

\usepackage[table, dvipsnames]{xcolor}
\usepackage{multirow}

\usepackage{calc}

\usepackage{enumitem}       

\usepackage{verbatim}
\usepackage{fancyvrb}

\usepackage{setspace}

\usepackage{graphicx, epsfig}

\usepackage{indentfirst}        